\documentclass[twocolumn]{aastex62}
\pdfoutput=1

\usepackage{amsmath}

\newcommand{\emin}{\epsilon_\mathrm{min}}
\newcommand{\temin}{\tilde{\epsilon}_{\mathrm{min}}}
\newcommand{\ellic}{\ell_\mathrm{IC}}
\newcommand{\ellgg}{\ell_{\gamma\gamma}}
\newcommand{\Bemin}{\tilde{B}\tilde{\epsilon}_\mathrm{min}}

\graphicspath{{./}{figures/}}
\allowdisplaybreaks

\shorttitle{Physics of Gaps in BH Magnetospheres II}
\shortauthors{Chen \& Yuan}

\begin{document}

\title{Physics of Pair Producing Gaps in Black Hole Magnetospheres II - General Relativity}

\correspondingauthor{Alexander Y. Chen}
\email{alexc@astro.princeton.edu}

\correspondingauthor{Yajie Yuan}
\email{yajiey@astro.princeton.edu}

\author{Alexander Y. Chen}
\affil{Department of Astrophysical Sciences, Princeton University, Princeton, NJ 08544, USA}

\author{Yajie Yuan}
\altaffiliation{Lyman Spitzer, Jr. Postdoctoral Fellow}
\affiliation{Department of Astrophysical Sciences, Princeton University, Princeton, NJ 08544, USA}

\begin{abstract}
    This is the second paper in a series where we examine the physics of pair
    producing gaps in low-luminosity accreting supermassive black hole systems.
    In this paper, we carry out time-dependent self-consistent fully general
    relativistic 1D PIC simulations of the gap, including full inverse Compton
    scattering and photon tracking. Similar to the previous paper, we find a
    highly time-dependent solution where a macroscopic vacuum gap can open
    quasi-periodically, producing bursts of $e^{\pm}$ pairs and high energy
    radiation. We present the light curve, particle and photon spectra from this
    process. Using an empirical scaling relation, we rescale the parameters to
    the inferred values at the base of the jet in M87, and find that the
    observed TeV flares could potentially be explained by this model under
    certain parameter assumptions.
\end{abstract}


\section{Introduction}
\label{sec:intro}
It is widely believed that powerful jets from Active Galactic Nuclei (AGN) are launched by rapidly spinning black holes through the Blandford-Znajek process \citep{1977MNRAS.179..433B}. When the black hole is threaded by a strong, coherent magnetic field, the rotating spacetime near the event horizon drags the field lines into rotation, resulting in a significant amount of Poynting flux flowing out along the magnetic field. The energy flow comes from the rotation of the black hole, and eventually gets dissipated further downstream along the jet, giving rise to the multi-wavelength emission we observe from jets. The launching and acceleration of Poynting flux dominated jets have been demonstrated in general relativistic magnetohydrodynamic (GRMHD) simulations \citep[e.g.,][]{2012MNRAS.423.3083M}. However, it turns out that in the jet funnel, plasma tends to be evacuated either inward to the black hole or outward to infinity, and some source of continuous plasma supply is needed to carry the underlying electric current, which cannot be self-consistently captured in the MHD framework.

In reality, when the plasma supply runs low in the jet funnel, electrostatic gaps --- regions with unscreened electric field --- could develop and accelerate particles to high enough energies to initiate an $e^\pm$ pair cascade, which would replenish the plasma needed in the jet circuit, similar to what happens around pulsars \citep[e.g.,][]{1977MNRAS.179..433B, 1992SvA....36..642B, 1998ApJ...497..563H}. But unlike pulsars where the pair producing photons are mainly generated by curvature radiation, near the black hole the dominant process is inverse Compton (IC) scattering of soft photons emitted by the accretion disk. The electrostatic gaps are probably most important in low luminosity AGN. In high luminosity AGN, the accretion disk may already produce many MeV photons that could collide with each other and convert to pairs in the jet funnel. If these pairs are already enough to conduct the current, then a gap does not need to open up \citep[e.g.,][]{2011ApJ...730..123L}.

The magnetospheric gap has been considered as possible candidate site for producing the rapid $\gamma$-ray variability seen from certain AGN, especially radio galaxies \citep[e.g.,][]{2011ApJ...730..123L, 2014Sci...346.1080A, 2017ApJ...841...61A, 2018ApJ...852..112K}. The most promising source seems to be M87 \citep{2018ApJ...852..112K}. It has a massive black hole with mass $M\approx6.5\times10^9M_{\odot}$ \citep{2019ApJ...875L...1E}, and an estimated jet power $L_{\rm jet}\sim 10^{44}\rm{erg}\,\rm{s}^{-1}$ \citep[e.g.,][]{2015ApJ...809...97B}. Its very high energy (VHE, $E>100$ GeV) $\gamma$-ray emission shows strong variability on timescales as short as a day \citep{2012ApJ...746..151A}. During the $\gamma$-ray flare, the VHE luminosity can reach $10^{42}\rm{erg}\,\rm{s}^{-1}$, and the flares have been observed to be accompanied by increased radio emission from the nucleus \citep{2009Sci...325..444A}, strongly suggesting particle acceleration in the vicinity of the black hole. In order to see whether the gap scenario can reproduce all the features of the $\gamma$-ray flares, a detailed study of the gap physics is needed.

So far in the literature there are quite a few works on the gap physics; most of them employ steady state/electrostatic models \citep{1992SvA....36..642B, 1998ApJ...497..563H,  2015ApJ...809...97B, 2016A&A...593A...8P, 2016ApJ...818...50H, 2016ApJ...833..142H, 2017ApJ...845...77H, 2018PhRvD..98f3016F, 2017PhRvD..96l3006L}. However, the gap development and screening may be an intrinsically time dependent process \citep{2015ApJ...809...97B, 2017PhRvD..96l3006L}, similar to the polar cap gap in pulsars \citep{2013MNRAS.429...20T}. Recently, \citet{2018A&A...616A.184L} used 1D general relativistic particle-in-cell (GRPIC) simulations to study the discharge of a gap that is depleted of particles at the beginning. They found that except for the initial transient, the long term evolution reaches a quasisteady state where no macroscopic gap is seen and pairs are generated everywhere randomly. \citet{2019PhRvL.122c5101P} carried out 2D GRPIC simulations of the jet launching process by immersing a rapidly spinning Kerr black hole in a uniform magnetic field. They used artificially simplified pair injection schemes, and found that the resulting plasma distribution depends sensitively on the pair creation physics.

In our previous paper, we also carried out 1D local simulations of the gap using a flat spacetime approximation \citep[hereafter CY18]{2018ApJ...863L..31C}. We found the gap development and screening to be a quasi-periodic process, a result qualitatively very different compared to \citet{2018A&A...616A.184L}. We focused on the high opacity scenario where inverse Compton scattering mainly happens in the Thomson regime, and constructed a simple analytic model to extrapolate numerical simulations to M87 and Sgr A*. It was found that for M87, the gap power is not enough to power the observed TeV flares. In this paper, we properly implement General Relativity and a more complete numerical treatment for inverse Compton scattering. We extend our study into the deep Klein-Nishina (KN) regime and attempt to find an empirical scaling relation to extrapolate our simulation results. We will first describe our method comprehensively in Section~\ref{sec:sim}, then present the main results in Section~\ref{sec:results}. We will comment on the implication of these results for M87 in particular in Section~\ref{sec:m87}, and finish with a discussion of potential caveats and future extensions in Section~\ref{sec:discussion}.

\section{Simulation setup}
\label{sec:sim}

\subsection{Equations of Motion}
\label{sec:eom}

We consider a nearly force-free magnetosphere around the black hole, with open field lines threading the event horizon, in a more or less axisymmetric, quasi-steady state. Somewhere along the field line, gaps may open up when particle density runs low. If the spatial and temporal scale of the gap is much smaller than the global characteristic scale ($\sim \text{a few } r_g\equiv GM/c^2$, the gravitational radius), the gap dynamics can be well described using 1D approximation along the field line.

In our numerical simulations, we take the deviation from the background force-free configuration to be our dynamic variable, and solve the equations along a 1D flux tube labeled by the flux function $\psi(r, \theta)$ (see Appendix \ref{sec:force-free} for a description of the force-free configuration). The basic equations have a relatively simple form in the $3+1$ formalism \citep[e.g.,][]{2004MNRAS.350..427K}. Here, the Kerr metric can be written as (we use Greek indices for 4-vectors, and Latin indices for spatial 3-vectors)
\begin{equation}
    ds^2=\left(\beta ^2-\alpha ^2\right)dt^2 +2\beta _i dt dx^i +\gamma _{ij}dx^i dx^j
\end{equation}
where $\alpha$ is the lapse function and $\beta_i$ is the shift vector. It is convenient to use Boyer-Lindquist coordinates as $\gamma_{ij}$ is diagnal: $\gamma_{ij}=\mathrm{diag}(g_{rr}, g_{\theta\theta},g_{\phi\phi})=\mathrm{diag}(\Sigma/\Delta, \Sigma, A \sin^2\theta/\Sigma)$, where $\Sigma=r^2+a^2\cos^2\theta$, $\Delta=r^2-2Mr+a^2$, $A=(r^2+a^2)^2-\Delta a^2\sin^2\theta$. Meanwhile,
\begin{align}
    \alpha&=\sqrt{\frac{g_{0\phi}^2-g_{00}g_{\phi\phi}}{g_{\phi\phi}}}=\sqrt{\frac{\Delta\Sigma}{A}},
\end{align}
and $\beta^i$ only has $\phi$ component:
\begin{equation}
    \beta^{\phi}=-\omega,
\end{equation}
where $\omega=-g_{0\phi}/g_{\phi\phi}=2Mar/A$ is the angular velocity of the zero angular momentum observer (ZAMO).
To alleviate the divergence near the event horizon, we use the tortoise coordinates which is a transformation from the Boyer-Lindquist coordinates by letting $d\xi = dr/\Delta$, such that
\begin{align}
    \xi (r)&=\frac{1}{r_+-r_-} \ln\left(\frac{r-r_+} {r-r_-}\right),\\
    r(\xi)&=\frac{r_+-r_-e^{\xi(r_+-r_-)}}{1-e^{\xi(r_+-r_-)}},
\end{align}
where $r_{\pm}=M\pm \sqrt{M^2-a^2}$ are the radius of the event horizons given by $\Delta=0$. We can see that $\xi\to-\infty$ as $r\to r_+$. The field equations that we solve in $(\xi, t)$ are:
\begin{align}
    \frac{\partial }{\partial \xi }(K_1 D^{\xi})=K_1 \left(\rho -\rho _{\rm ff}\right),\label{eq:gauss}\\
    -\frac{\partial }{\partial t}(K_1 D^{\xi })=K_1 \left(j^{\xi }-j_{\rm ff}^{\xi }\right),\\
    \frac{\partial }{\partial t}(K_1 \rho) +\frac{\partial }{\partial \xi }(K_1 j^{\xi })=0,
\end{align}
where the factor $K_1\equiv\Delta\sqrt{\gamma}/(\partial_{\theta} \psi)$ accounts for field line divergence along the flux tube. We use the third equation to solve for the current due to particle movement, therefore Gauss's law (equation \eqref{eq:gauss}) is automatically satisfied as long as it holds initially. We always set $\rho = \rho_{\rm ff}$ initially and $D^{\xi} = 0$, which trivially satisfies the Gauss's law. We observe excellent charge conservation as even after millions of time steps charge density is always driven back to $\rho_{\rm ff}$.

For particles, we assume they are well magnetized and can only move freely along magnetic field lines. In other words, their motion is constrained to 1D, and $u^{\xi}=u^r/\Delta$ completely determines the full 4-velocity:
\begin{align}
    u^{\phi}&=\Omega u^0+\frac{B^{\phi}}{B^{\xi}}u^{\xi},\\
    u^{\theta}&=\frac{B^{\theta}}{B^{\xi}}u^{\xi},
\end{align}
where $B^{\xi}=B^r/\Delta$, and $\Omega$ is the angular velocity of the field line. We can then write down the Lagrangian in terms of coordinate $\xi$ and velocity $v_{\xi}=d\xi/dt$:
\begin{equation}
    L(t,\xi,v^{\xi})=-\frac{1 }{u^0}-\frac{q }{m}\Phi,
\end{equation}
where $u^0$ is written in terms of $v^{\xi}$ as
\begin{equation}
    u^0=\frac{1}{\sqrt{\alpha ^2-\left(S_2\left( v^{\xi}\right)^2+2 S_1 v^{\xi}+S_3\right)}},
\end{equation}
here
\begin{align*}
  S_1&=(\beta_{\phi}+\Omega\gamma_{\phi\phi})\frac{B^{\phi}}{B^{\xi}},\\
  S_2&=\gamma_{\xi\xi}+\gamma_{\phi\phi}\left(\frac{B^{\phi}}{B^{\xi}}\right)^2+\gamma_{\theta\theta}\left(\frac{B^{\theta}}{B^{\xi}}\right)^2,\\
  S_3&=\beta^2+2\Omega\beta_{\phi}+\gamma_{\phi\phi}\Omega^2.
\end{align*}
The canonical momentum is
\begin{equation}\label{eq:p_xi}
    p_{\xi}=\frac{\partial L}{\partial v^{\xi}}=u^0(S_1+S_2 v^{\xi}),
\end{equation}
and the Euler-Lagrange equations are
\begin{align}\label{eq:1D_particle_EoM}
    \frac{dp_{\xi}}{dt}=\frac{\partial L}{\partial \xi}&=\frac{q }{m}E_{\xi}-\frac{u^0}{2}\frac{\partial }{\partial \xi}\alpha ^2\nonumber\\
    &+\frac{u^0}{2}\left(\frac{\partial S_3}{\partial \xi}+2\frac{\partial S_1}{\partial \xi}v^{\xi}+\frac{\partial S_2}{\partial \xi}\left(v^{\xi}\right)^2\right),
\end{align}
where $E_{\xi}=-\partial_{\xi}\Phi$.
The equations can also be written purely in terms of $p_{\xi}$, using
\begin{align}
    v^{\xi}&=\frac{p_{\xi}/u^0-S_1}{S_2},\label{eq:v_xi}\\
    u^0&=\sqrt{\frac{p_{\xi}^2+S_2 }{S_2(\alpha ^2-S_3)+S_1^2}}.\label{eq:u0-in-p_xi}
\end{align}
In our field update equations $D^{\xi}$ is the dynamic variable, while Equation (\ref{eq:1D_particle_EoM}) needs $E_{\xi}$. We can obtain the latter through
\begin{equation}
    E_{\xi}=\alpha D_{\xi}=\alpha\gamma_{\xi\xi}D^{\xi}.
\end{equation}
This set of equations naturally reproduces the inner and outer light surfaces, outside which particles flow either to the horizon or to infinity. We place the boundaries of our simulation box to be slightly outside the two light surfaces, so that all particles near the boundary can only outflow.

Photons are not confined to field lines so they obey the general equations of motion.
In our example, the background field lines on the poloidal plane is more or less radial. It is a good approximation to neglect $\theta$ component of the photon 4 velocity. Therefore the equations are
\begin{align}
    \frac{du_{\xi}}{dt}&=- \frac{u^0}{2}\partial_{\xi} \alpha^2+u_{\phi }\,\partial_{\xi} \beta ^{\phi }-\frac{u_{\xi}^2\partial_{\xi} \gamma^{\xi\xi}+u_{\phi}^2\,\partial_{\xi}\gamma^{\phi\phi}}{2 u^0},\\
    \frac{du_{\phi}}{dt}&=0,\\
    \frac{d\xi}{dt}&=\gamma ^{\xi\xi}\frac{u_{\xi}}{u^0},\\
    u^0&=\frac{1}{\alpha }\sqrt{\gamma ^{\xi\xi}\left(u_{\xi}\right)^2+\gamma ^{\phi \phi }\left(u_{\phi }\right)^2},
\end{align}
where $\gamma^{\xi\xi}=\gamma^{rr}/\Delta^2$.

\subsection{Radiative Transfer}
\label{sec:rad}

We assume that the background soft photon field is isotropic in the ZAMO frame, and model the scattering event (IC and $\gamma\gamma$ pair production) in this frame. For simplicity, we assume that the number density of the soft photons in the ZAMO frame is a power law independent of the radius:
\begin{equation}
    n(\epsilon) = \frac{n_0}{\epsilon_{\min}}\left(\frac{\epsilon}{\emin}\right)^{\alpha-1}
\end{equation}
We have chosen the normalization such that $n_0$ has the dimension of the total number of photons per volume.
In addition, we consider the regime where the number density of the background soft photon field is much larger than those emitted by the particles in the gap, so that only scatterings on the background photons are important while nonlinear effects can be neglected.

The rate of scattering between an electron and the background photons in the ZAMO frame is \citep[e.g.,][]{Blumenthal1970RvMP...42..237B}
\begin{equation}
\hat{\nu}_{\rm IC}=\int \frac{1}{2} n(\epsilon)\sigma _{\text{IC},\epsilon } c  (1-\beta  \cos \theta  ) \sin\theta\, d\theta\,d\epsilon.
\end{equation}
Here $\sigma _{\text{IC},\epsilon }$ is the full KN cross section \citep[e.g.,][]{rybicki_radiative_1986}:
\begin{equation}
\begin{split}
    \sigma _{\text{IC},\epsilon }=\sigma _T \frac{3}{4}\left[\frac{(x+1)}{x^3} \left(\frac{2 x (x+1)}{2 x+1}-\ln(2 x+1)  \right)\right.\\
    \left.+\frac{\ln (2 x+1) }{2 x}-\frac{3 x+1}{(2 x+1)^2}\right],
\end{split}
\end{equation}
where $x$ is the incident photon energy $\epsilon'$ in the electron rest frame measured in units of the electron rest mass:
\begin{equation}
x=\frac{\epsilon'}{m_e c^2}=\frac{\gamma  \epsilon  (1-\beta   \cos \theta )}{m_e c^2},
\end{equation}
and $\gamma=\alpha u^0$ is the energy of the electron in the ZAMO frame. At each time step, we determine the scattering probability $P_{\rm IC}$ using the following relation:
\begin{equation}
    P_{\rm IC}=\hat{\nu}_{\rm IC}\Delta \hat{t}=\hat{\nu}_{\rm IC}\alpha\Delta t,
\end{equation}
where $\Delta t$ is the global time step, and $\Delta \hat{t}$ its corresponding value in ZAMO frame. If a scattering event occurs, we draw the scattered photon energy based on the IC spectrum  \citep[e.g.,][]{1968PhRv..167.1159J}
\begin{equation}
\begin{split}
    \frac{dN}{dE_1 d\hat{t}}&=\int_{\epsilon_{\rm min}}^{\epsilon_{\rm max}}\frac{n(\epsilon) d\epsilon }{\epsilon }\frac{2 \pi  r_e^2 m_e c^3}{\gamma }\times\\
    &\left[2 q \ln q+(1+2 q)(1-q) +\frac{1}{2}\frac{ \left(\Gamma _{\epsilon }q \right)^2}{ \left(q \Gamma _{\epsilon }+1\right)}(1-q) \right],
\end{split}
\end{equation}
where $E_1$ is the energy of the scattered photon in units of the initial electron energy: $\epsilon_1=\gamma m_e c^2 E_1$; $\Gamma_{\epsilon}=4 \epsilon \gamma/(m_e c^2)$, $q=E_1/[\Gamma_{\epsilon}(1-E_1)]$. The range of $E_1$ is
\begin{equation}
1\gg\frac{\epsilon }{\gamma  m_e c^2}\leq E_1\leq \frac{\Gamma _{\epsilon }}{\Gamma _{\epsilon }+1}.
\end{equation}
Since the photon is not tied to the field line, we compute its $u_\xi$ and $u_\phi$ when it is created, and follow its trajectory using the photon equations of motion in section \ref{sec:eom}.

For $\gamma\gamma$ pair production, the total cross section for photons of energy $E$, $\epsilon$ and incident angle $\theta$ is \citep{Gould1967}
\begin{equation}
\sigma_{\gamma\gamma} =\frac{1}{2} \pi r_e^2 \left(1-\beta ^2\right) \left[\left(3-\beta ^4\right) \ln \frac{1+\beta}{1-\beta }-2 \beta  \left(2-\beta ^2\right)\right],
\end{equation}
where $\beta c$ is the electron  (positron) velocity in the center of mass frame,
\begin{equation}
\beta=\left(1-\frac{1}{s}\right)^{1/2},
\quad \text{with }
s=\frac{E \epsilon  (1- \cos\theta  )}{2m_e^2 c^4 }.
\end{equation}
The threshold requires $s>1$. A high energy photon of energy $E$ traversing an isotropic background soft photon field with a number spectrum $n(\epsilon)$ in the ZAMO frame will have an absorption probability per unit time
\begin{equation}
    \hat{\nu}_{\gamma\gamma}=\int \frac{1}{2}n(\epsilon)\sigma_{\gamma\gamma}c(1-\cos\theta)\sin\theta\,d\theta\,d\epsilon.
\end{equation}
When a photon converts to a pair, we set the direction of motion of the particles to be along local $B$ field, and the photon energy is evenly split to the pair in the local ZAMO frame.

We tabulate all the cross sections and spectra and use a binary search table look-up to find the correct rates. This is an efficient procedure and we are able to process more than $10^9$ scatterings per second on a single GPU.

\subsection{Units and Scales}
\label{sec:scales}

We adopt a numerical unit system where $r_g$ is the unit of length, and $r_g/c$ is the unit of time. Magnetic field in numerical units is measured in terms of cyclotron frequency. We use tilde to denote dimensionless quantities. For example, $\tilde{B} = eBr_g/(m_e c^2)$.

There are three main parameters of the problem. $B_0$ is the background magnetic field strength, which determines the background current $j_{\rm ff}$ and charge density $\rho_{\rm ff}$. It also sets a characteristic plasma skin depth
\begin{equation}
    \lambda_p = \frac{c}{\omega_p} = \sqrt{\frac{m_e c^3}{4\pi j_{\rm ff} e}} \sim \sqrt{\frac{m_e c^2r_g}{B_0e}},
\end{equation}
which leads to $\tilde{B}_0 = (r_g/\lambda_p)^2$. In other words, $\tilde{B}_0$ sets the ratio between macroscopic and microscopic length scales.

The peak of the soft photon spectrum $\emin$ is the second parameter, which sets the energy scale of the radiative transfer. Particles with $\gamma \lesssim 0.1/\temin$ interact with the soft photon field in the Thomson regime, while those with $\gamma > 0.1/\temin$ fall into the KN regime and suffer from reduced scattering cross section. Gamma-ray photons with energy $0.1/\temin \lesssim\tilde{\epsilon} \lesssim 1/\temin$ have the smallest free path to $\gamma$-$\gamma$ collision.

\begin{figure*}[t]
    \centering
    \includegraphics[width=\textwidth]{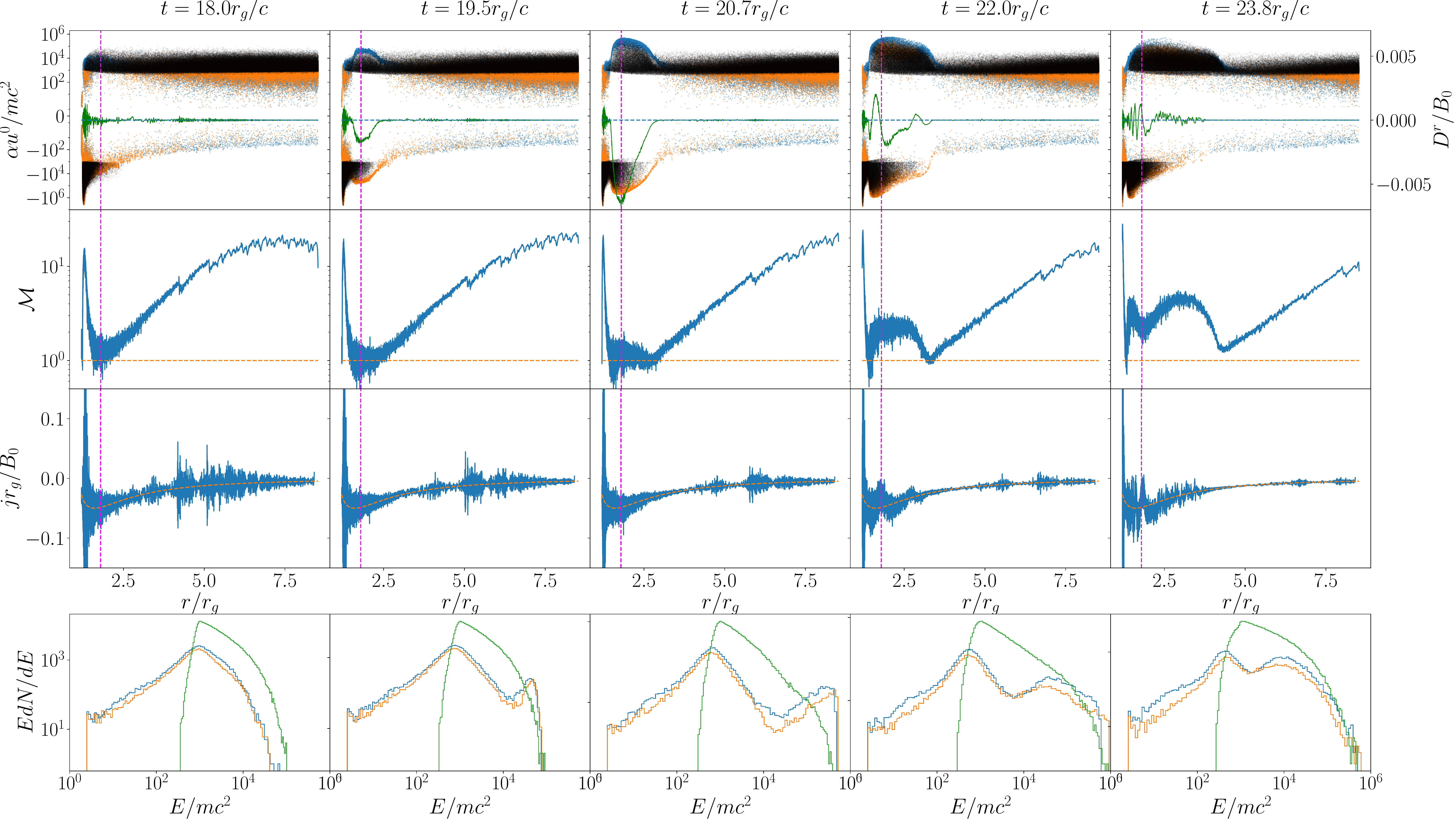}
    \caption{Time evolution of the gap. From left to right are snapshots at
      labeled times. The 4 panels from top to bottom are: 1) Phase space plots
      for electrons (blue), positrons (orange), and photons (black). The green
      line is electric field and its scale is on the right. 2) Pair multiplicity
      $\mathcal{M}$ defined in equation \eqref{eq:multiplicity}. Orange dashed
      line marks $\mathcal{M} = 1$. 3) Current $j$ in blue and its background
      value $j_\mathrm{ff}$ in orange. 4) Spectrum of electrons (blue),
      positrons (orange), and photons (green) in the box, excluding particles
      very close to the inner light surface. The vertical dashed line in the
      first three rows indicates the null surface. The parameters for this run
      are: $\tau_0 = 10$, $\tilde{B}_0 = 10^8$, and $\temin = 10^{-5}$. The
      simulation box has 196608 grid points.}
    \label{fig:gap-evol}
\end{figure*}

The third parameter $\tau_0$ is the characteristic optical depth to inverse Compton scattering in the Thomson regime. It scales with the number density $n_0$ of the soft photon field:
\begin{equation}
    \tau_0 = r_gn_0\sigma_T = \frac{r_g}{\ell_{\rm IC}}=(\tilde{\ell}_{\rm IC})^{-1}
\end{equation}
where $\ell_{\rm IC}=1/(n_0\sigma_T)$. $\tau_0$ determines the efficiency of IC scattering as well as $\gamma$-$\gamma$ collision. If any gap develops in the magnetosphere, it is difficult to screen it before it grows to $h \sim r_g/\tau_0$, simply because that is the minimum mean free path the high energy photons need to travel before creating pairs. If $\tau_0$ is close to unity, then we expect the gap to at least grow to scale of $r_g$.

We use the code \emph{Aperture} for simulations in this paper. This version of the code is highly optimized on GPUs, and provides a unified framework for simulating the magnetospheres of neutron stars and black holes. The resolution for the runs in this paper range from 16384 to 196608 grid cells, with up to $200$ million particles and photons, evolved over several million time steps. All of these runs were carried out on a single GPU, either a Tesla P100 or a Quadro P6000.

\section{Results}
\label{sec:results}

\subsection{Time-dependent Gap}
\label{sec:gap}

\begin{figure*}[t]
    \centering
    \plottwo{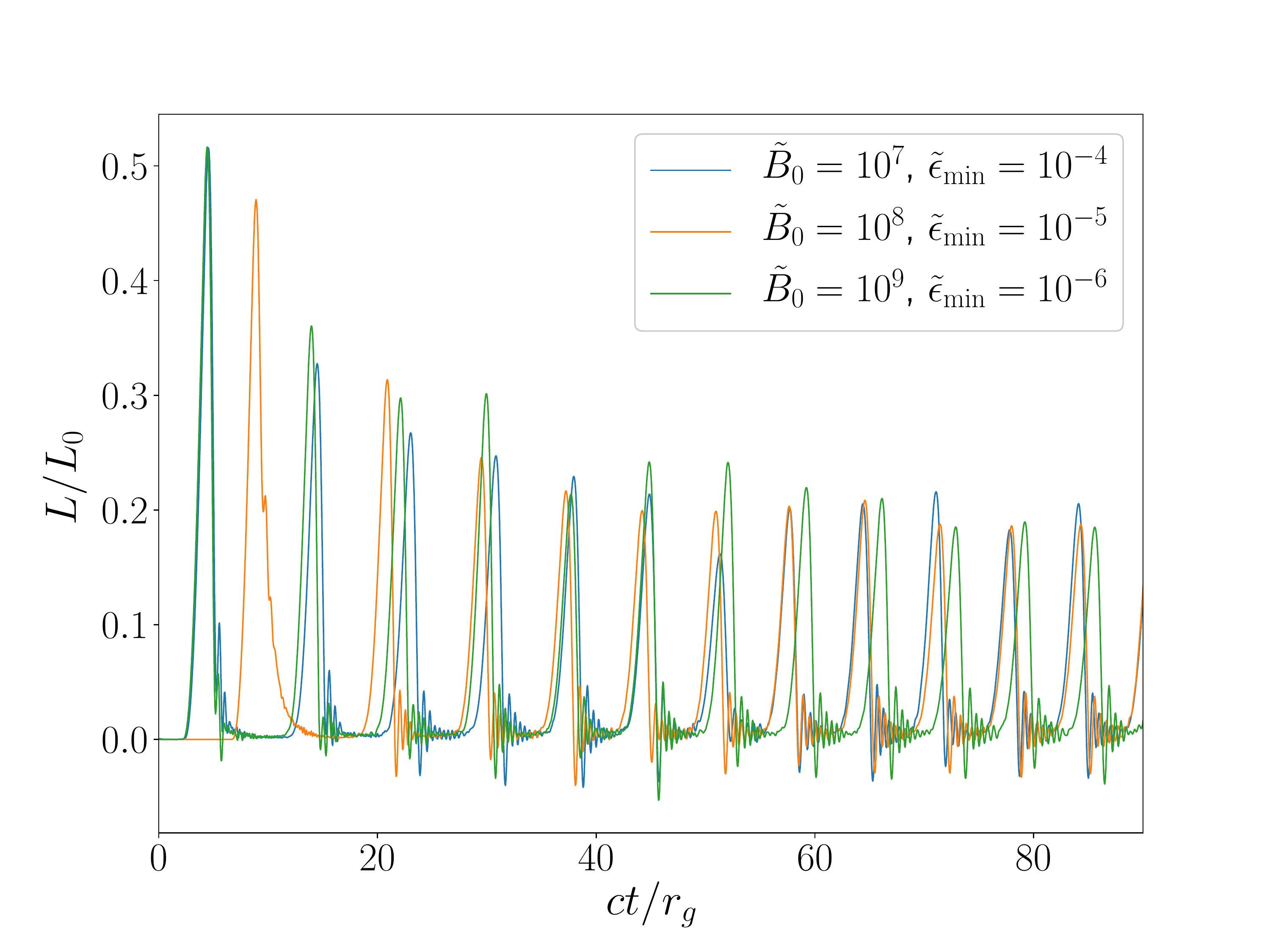}{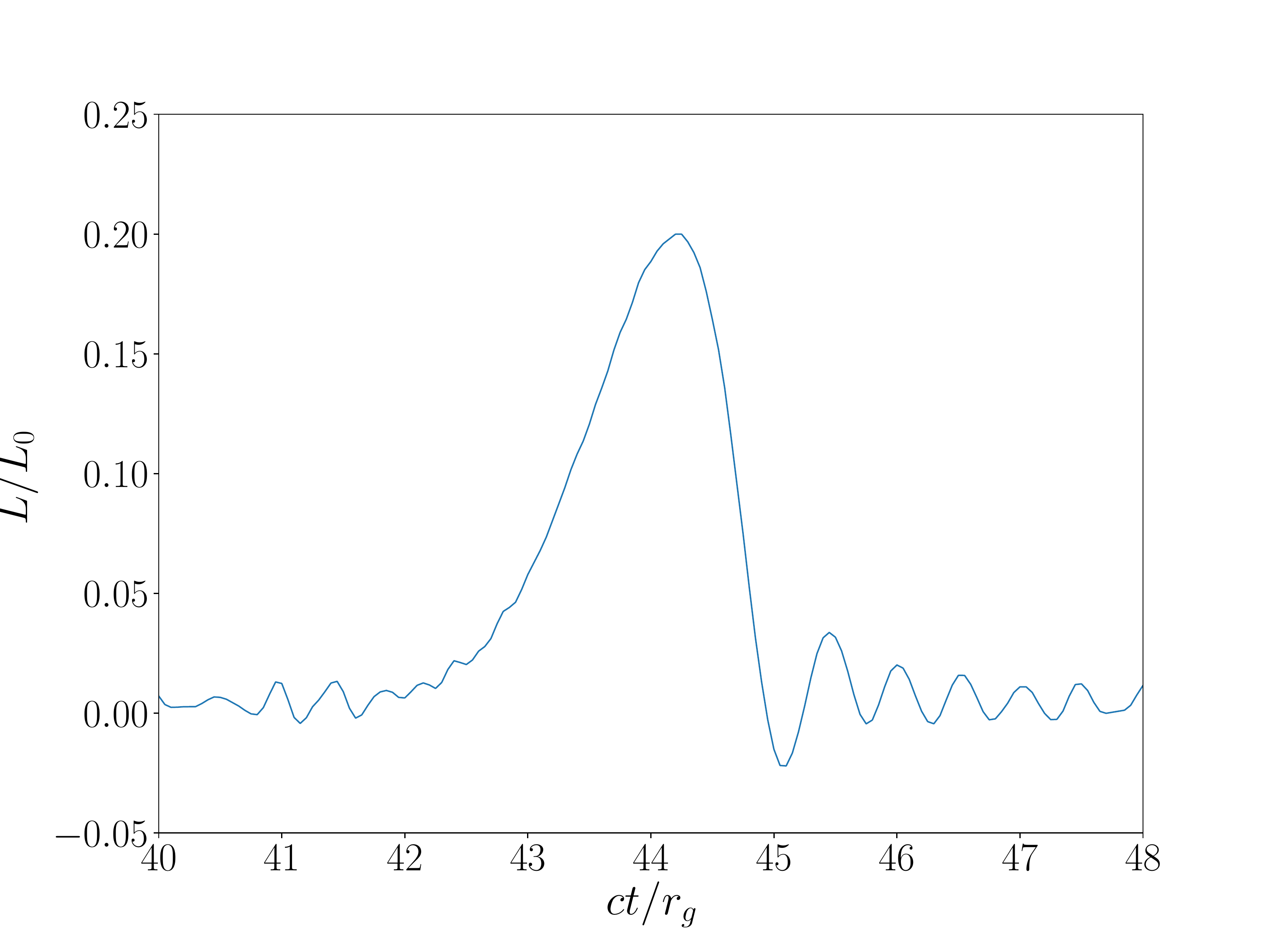}
    \caption{Left panel: Dissipation rate as a function of time, for 3 different simulations with the same $\tau_{0} = 10$, normalized to the jet power. The three runs, despite having orders of magnitude difference in $\tilde{B}_{0}$ and $\temin$, exhibit roughly the same gap screening and recurrence time scales, as well as similar amount of overall dissipation. The second run (orange line) has a different initial condition with higher multiplicity, hence a different profile of initial transient, but it settles down to the same quasi-periodic behavior. Right panel: Zoomed-in light curve for the orange line in the left panel to show the shape of a single burst.}
    \label{fig:light-curves}
\end{figure*}

In almost all cases, our simulations show quasi-periodic gap opening and screening, very similar to the results of CY18 
. Typically the system will undergo an initial transient phase where a large gap opens, filling the box with plasma, then settle down to the quasi-periodic state. One such cycle is shown in Figure~\ref{fig:gap-evol}. Here a good criterion for forming a gap is based on the multiplicity, defined as the ratio between the actual number of charge carriers in the gap and the minimum amount needed to provide the current\footnote{In ZAMO frame, the charge density is $\hat{\rho}=\rho$, while the radial current density is $j^{\hat{r}}=j^r\sqrt{g_{rr}}/\alpha$, so $\mathcal{M}=|\hat{\rho}_+-\hat{\rho}_-|c/j^{\hat{r}}$.}:
\begin{equation}
    \mathcal{M} = \frac{|\rho_{+}-\rho_{-}|\alpha c}{\sqrt{g_{rr}}j_\mathrm{ff}^r}.
    \label{eq:multiplicity}
\end{equation}
Whenever the local pair multiplicity drops below unity, parallel electric field starts to grow, accelerating particles and producing photons. It is remarkable that, except for the initial transient which depends on initial condition, the gap always grows from the null surface where $\rho_{\rm ff} = 0$.

The quasi-periodic behavior of the gap can be seen from the integrated dissipation rate in the flux tube, $L = \int \alpha g_{rr}D^r j^r\,(\sqrt{\gamma}/\partial_{\theta}\psi)dr$.
The time-evolution of this gap power can be interpreted as the light curve, and is plotted in Figure~\ref{fig:light-curves}, where the power is normalized to the Poynting flux density in the same flux tube measured in the force-free simulation, $L_0 \approx 0.0124B_0^2$, which we can also identify as the jet power. Three runs of different $\tilde{B}_0$ and $\temin$ but same $\tilde{B}_0\temin$ and $\tau_0$ are plotted, and they have near identical time evolution and gap dissipation. As will be discussed in section~\ref{sec:parameters}, we find that in general gap power and screening timescale only scale with the product $\tilde{B}_0\temin$, but insensitive to the values of $\tilde{B}_0$ and $\temin$ themselves. 

\begin{figure}[h]
    \centering
    \plotone{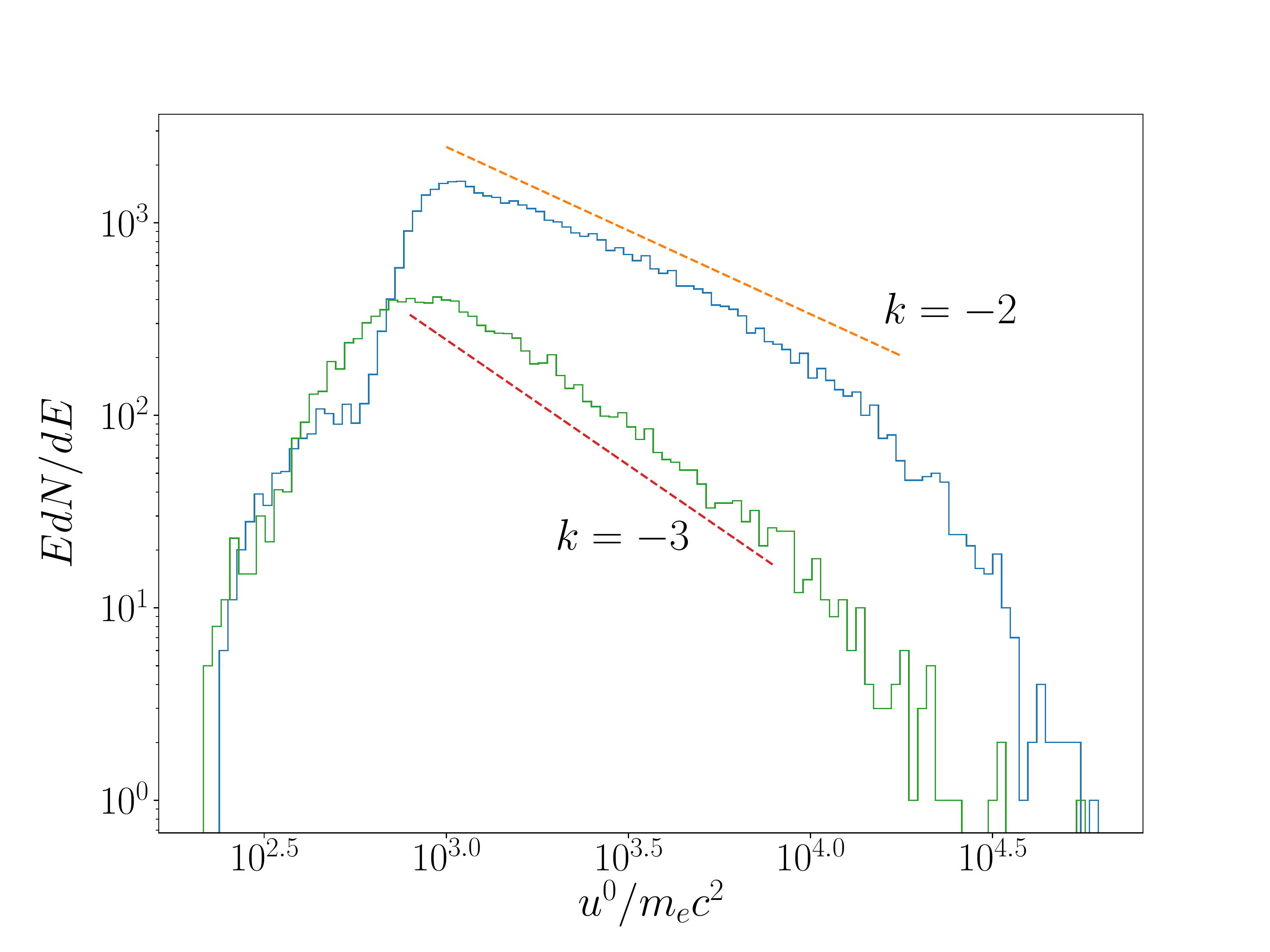}
    \caption{Outgoing photon spectrum at high (blue curve) and low (green curve) states defined in section \ref{sec:gap}. $k$ is the slop of the power law index in $EdN/dE$. The low energy cutoff is mostly due to our removal of photons with energies below $\sim 10^{3}m_{e}c^2$ at creation. This particular run has $\tilde{B}_0 = 10^8$ and $\temin = 10^{-5}$.}
    \label{fig:spectra}
\end{figure}

In addition to highly time-dependent gap power, we observe the same variability reflected in the photon flux near the outer light surface. We keep track of about 5\% of all the particles and photons in the simulation, and look at all the photons within $0.5r_g$ of the outer light surface. The total energy as well as the spectrum of these outgoing photons vary according to gap activity. Figure \ref{fig:spectra} compares the spectra of the high state (when photon flux is highest) and low state (when photon flux is lowest). Both spectra form a power law, with the high state having a harder spectral index. The photon spectrum tends to extend a bit above $0.1/\temin$, at which point it is strongly absorbed. The lower end of the spectrum is somewhat artificial, since we need to remove some low energy photons in the simulation in order to make it computationally feasible. In this particular run, all photons with energy below $10^{3}m_{e}c^2$ are removed at creation. However, photons created above this energy threshold are allowed to propagate and can be redshifted to lower energies if they travel from near the gap to the outer boundary. Even with lower energy photons removed, the energy escaping from the box is still dominated by photons, suggesting most of the gap power goes into radiation. 

In our simulations the multiplicity of pairs escaping the box is usually between 10--50. However, most of the photons emitted in the gap do not convert in the box, and will continue to produce more particles if they travel outwards for a much larger distance, undergoing a post gap cascade which has been discussed by e.g. \citet{2015ApJ...809...97B}. We expect the terminal multiplicity to be much higher than what we get in the local simulation.

\begin{figure*}[t]
    \centering
    \plottwo{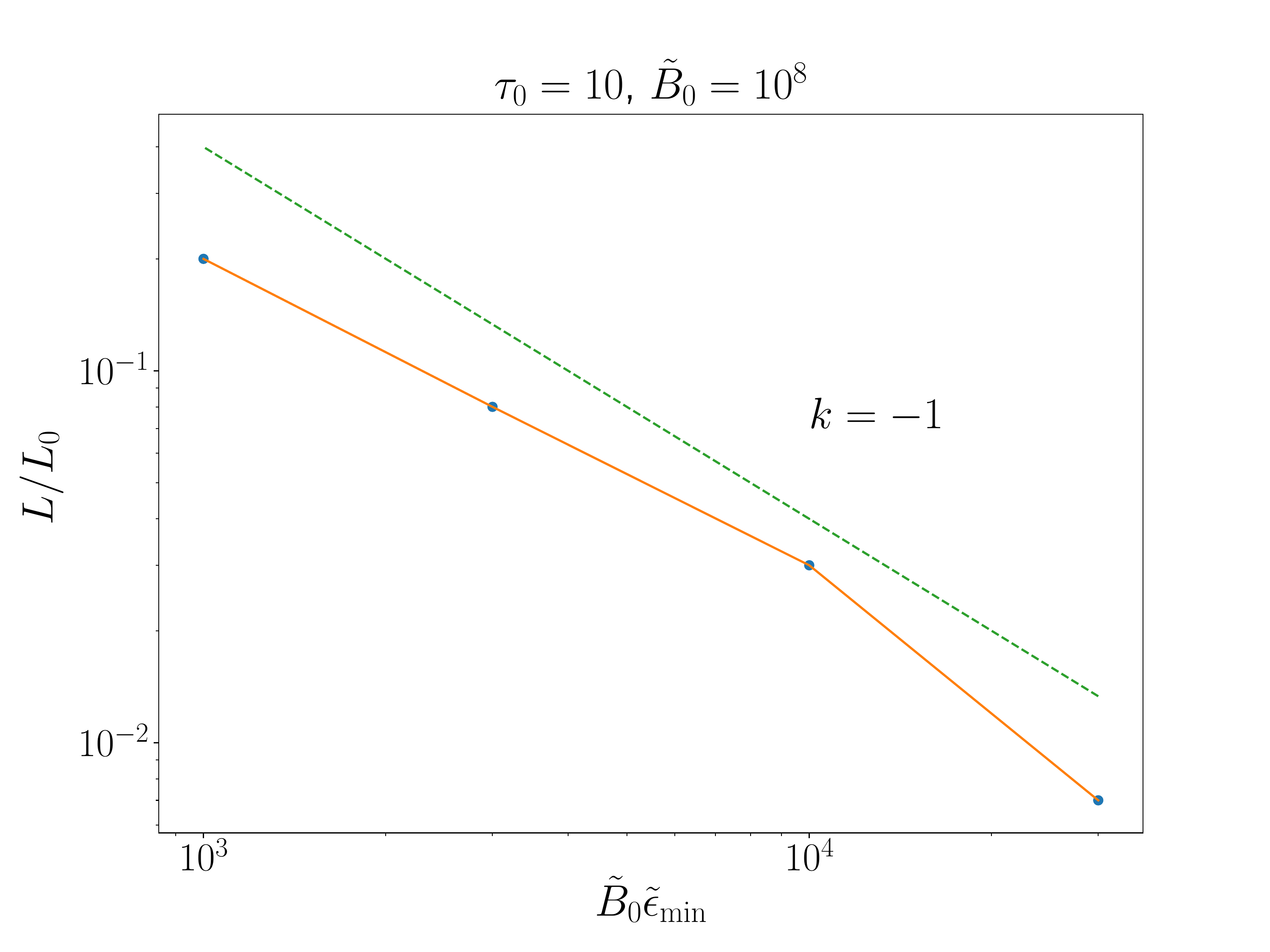}{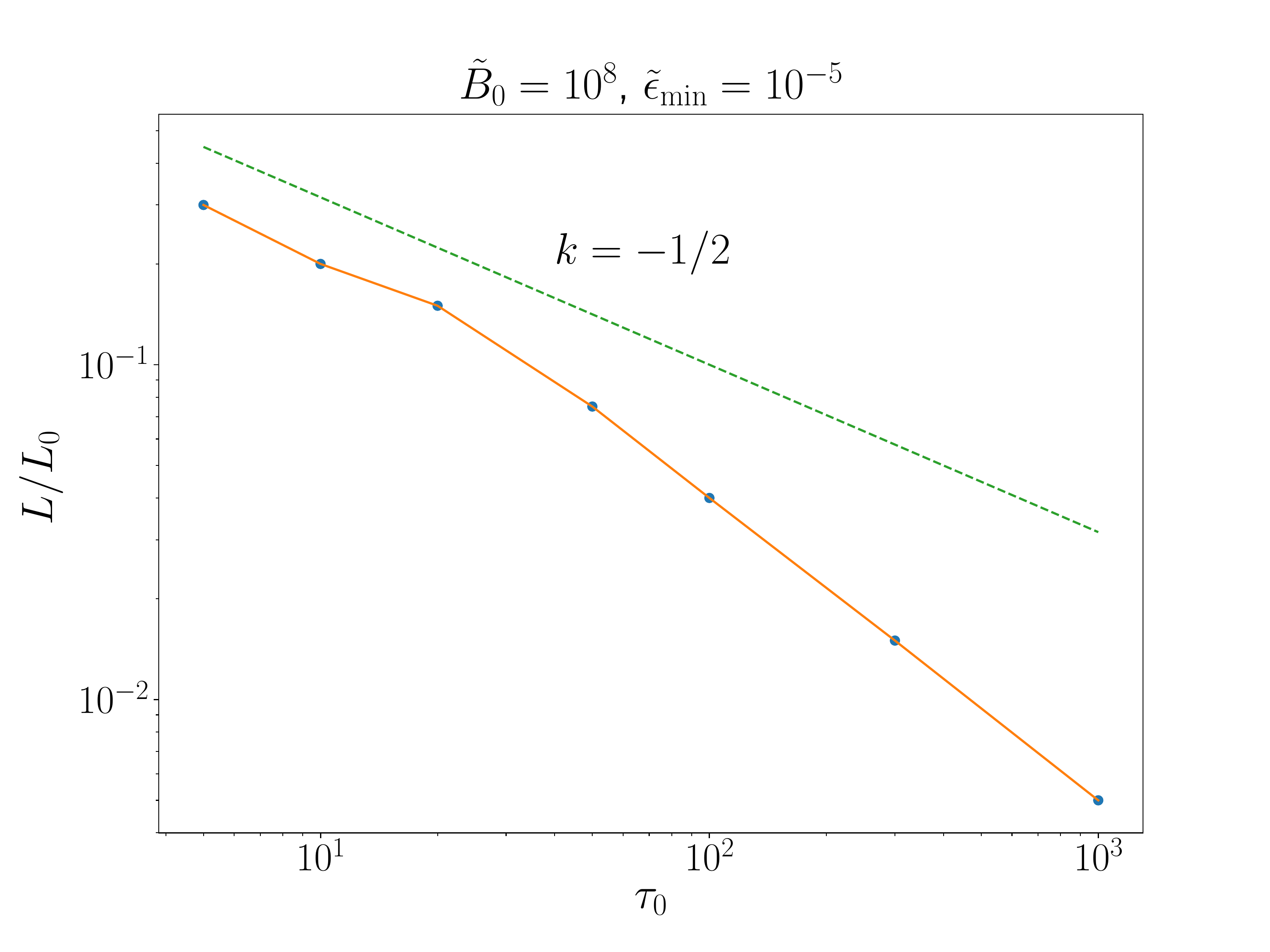}
    \caption{Scaling of the gap power with respect to $\tilde{B}_0\temin$ (left) and $\tau_0$ (right). The green dashed lines are added to illustrate potential scaling relationship. The luminosities $L$ are measured at the peak averaged over several cycles.}
    \label{fig:scaling}
\end{figure*}

\subsection{Dependence on Parameters}
\label{sec:parameters}

In the high $\tau_0$ or small $\ellic$ case, e.g. $\tau_0 \gtrsim 100$, IC scatterings mainly happen in the Thomson regime. The behavior of the gap is identical to what was reported in
CY18. Despite the high optical depth, the gap is macroscopic with a size $h \sim 0.1\text{-}1r_g$, sometimes even larger during the initial transient. Particles are predominantly radiation reaction limited, and gap screening happens when energies of the upscattered photons are increased such that $\ellgg \sim h$. The fiducial optical depth, together with $\temin$, sets the energy scale at which gap screening occurs. The scaling relation derived in
CY18, namely their equation (15), still holds in the high $\tau_0$ regime. Using the unit convention of this paper, the same relation can be written as (neglecting the order 1 constants)
\begin{equation}
  \frac{\tilde{D}_{\max}^{\hat{r}}}{\tilde{B}_0}\sim\left(\tilde{B}_0\temin\right)^{-\frac{\alpha}{\alpha+1}}\tau_0^{-\frac{2-\alpha}{\alpha+1}}.
\end{equation}
Notice that this ratio depends only on the product of $\tilde{B}_0$ and $\temin$, not individually. With fixed $\tau_0$ the gap size is also almost constant with respect to the product $\tilde{B}_0\temin$. Although this equation should only be valid in the Thomson regime, we find that the dependence on the product $\Bemin$ is more universal. This is because particle energy only has one characteristic scale, which is $\gamma_0=m_e c^2/\epsilon_{\min}$. Writing $\gamma=\tilde{\gamma}\gamma_0$, electric field $E=\tilde{E}B_0$, particle number density $n=\tilde{n}m_e c^2/(e^2r_g^2)$, the main equations become (neglecting the geometric factors) $\partial_{\tilde{r}}\tilde{E}\sim \tilde{n}_+-\tilde{n}_+$, and $\partial_{\tilde{t}}\tilde{\gamma}\sim (\tilde{B}_0\tilde{\epsilon}_{\min})\tilde{E}$, which only depend on the product $\tilde{B}_0\temin$.

When $\tau_0 \lesssim 100$, particles are accelerated into the KN regime. Since the photon mean free path for $\gamma$-$\gamma$ collision is smallest when the photon energy is comparable to $m_e c^2/\temin$, the gap will be at least as large as the minimal $\ellgg$, which is $\ellgg \sim 5\ellic = 5r_{g}/\tau_0$. We observe that the gap size scales very weakly with parameters, only becoming larger when $\tilde{B}_0\temin$ is so small that the gap voltage $ehD_\mathrm{max}^{\hat{r}}$ is comparable to $m_c c^{2}/\temin$. Surprisingly, when $\tau_0 \lesssim 3$, the particles are accelerated too far into the KN regime that the gap is no longer screened in our simulations.

The gap screening mechanism in the KN regime is different from what was described in CY18 for the Thomson regime. In the latter case IC loss is efficient so that most of the highest energy particles in the gap are radiation reaction limited, and the highest energy photons have the smallest free path. It is the highest energy photons that eventually produce the gap-screening pairs, therefore we focused on the kinematics of the highest energy particles to derive the gap scaling relation. However, in the KN regime this is no longer the case. Cooling becomes inefficient when the particle Lorentz factor reaches $1/\temin$. Particles are accelerated to much higher energies than the radiation reaction limit, and the first secondary pairs are created by photons with $\epsilon_\gamma \sim m_e c^2/\temin$, since they have the shortest mean free paths. However, $\ellgg$ is large even for these photons, so there is a significant delay between the emission of these photons and the screening of the gap, during which time the primary particles are accelerated to even higher energies, e.g. $\gamma\gtrsim 10/\temin$. The screening process in the KN regime depends strongly on the acceleration history of the primary particles in the gap. When $\ellgg\gtrsim r_g$, the gap is global and the electric field grows so quickly that not enough photons with energy $\sim m_ec^2/\temin$ are emitted. As a result, the gap can no longer be screened. 

\begin{figure*}[t]
    \centering
    \plottwo{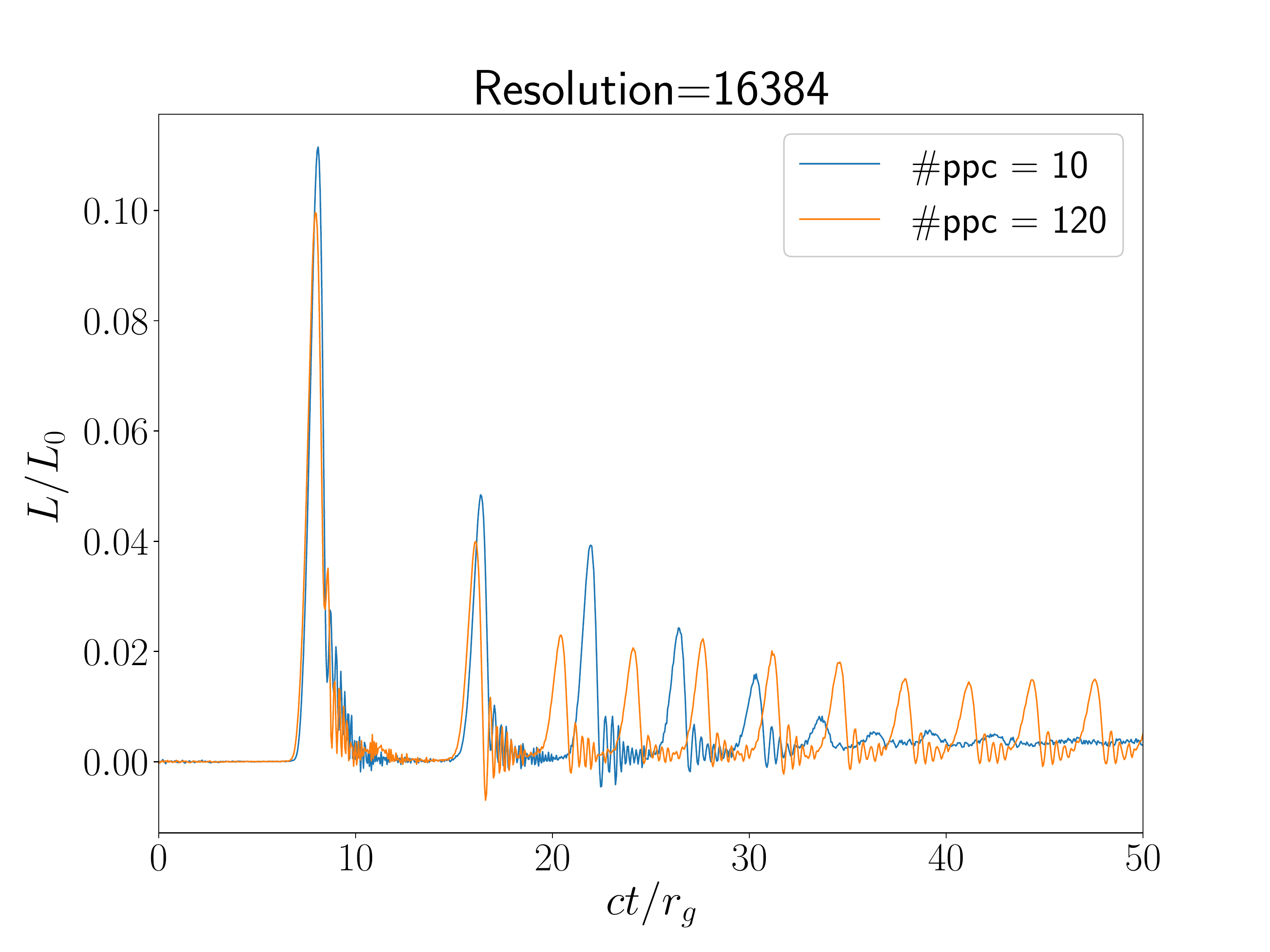}{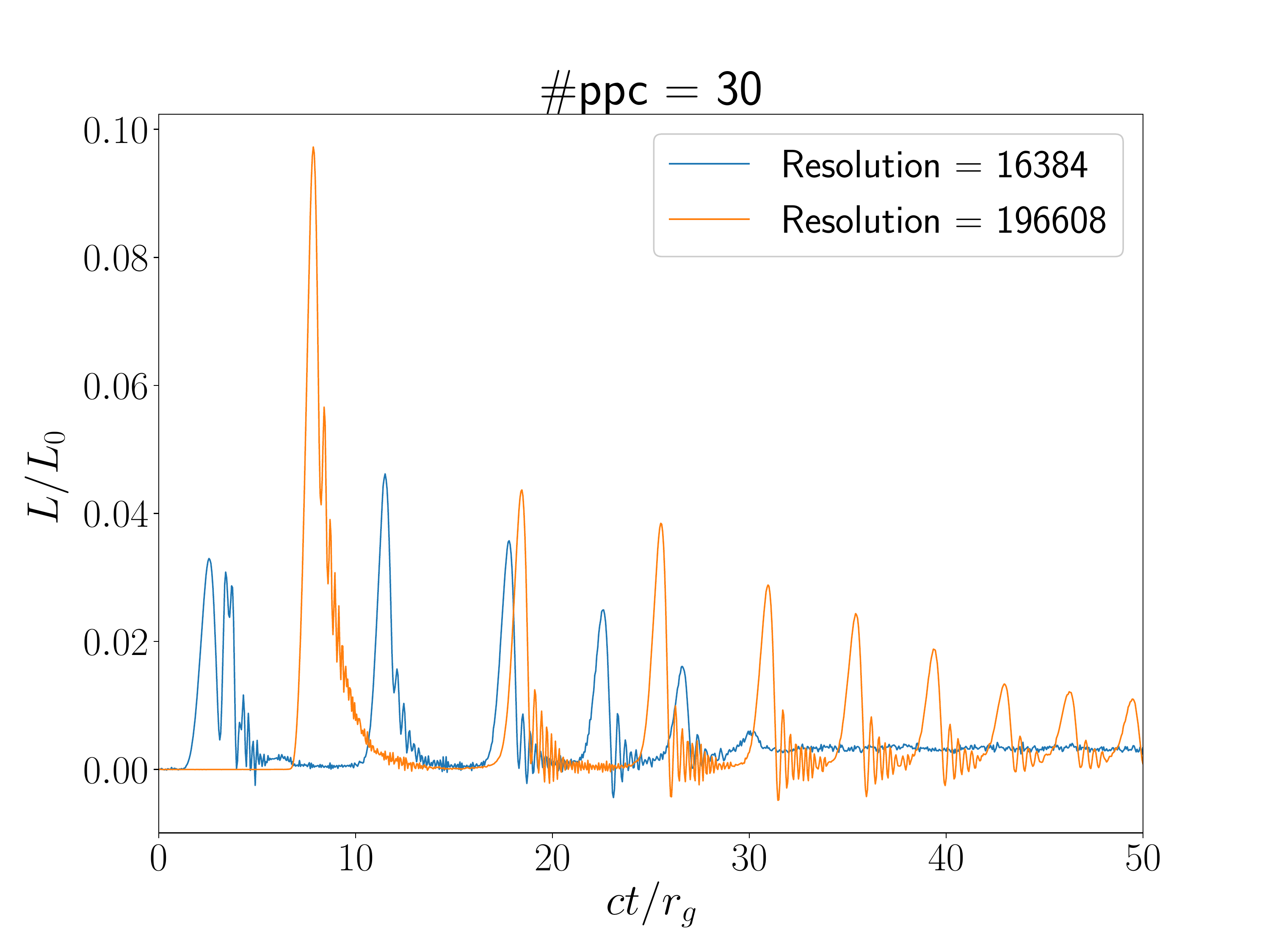}
    \caption{Quasi-steady state vs. time-dependent gap for different numerical parameters. Left panel: Light curves of two simulations with $\tilde{B}_0 = 10^7$, $\temin = 10^{-3}$, and $\tau_0 = 10$, but different initial particles per cell. Right panel: Light curves of two simulations with $\tilde{B}_0 = 10^8$, $\temin = 10^{-4}$, and $\tau_0 = 10$, but different grid resolution and slightly different initial conditions (hence a different initial transient). Here ppc means the number of electrons and positrons per cell in the initial condition where $\rho = \rho_\mathrm{ff}$}.
    \label{fig:resolution}
\end{figure*}

Figure~\ref{fig:scaling} shows the dependence of the gap power on $\Bemin$ and $\tau_0$ individually. In all the runs shown on the plot, we observe the same time-dependent gap described in~\ref{sec:gap}, and measure the  peak gap power averaged over several cycles. It can be seen that in general the power scales close to $L\sim (\Bemin)^{-1}$ with fixed $\tau_0$. However, we have to note here that at high $\Bemin$ the measured gap power could be smaller than it should be, due to numerical heating which will be discussed in section \ref{sec:resolution}, since numerical heating tends to reduce the need for gap acceleration.

Also in Figure~\ref{fig:scaling} one can see that when $\tau_0$ is around $10$ the gap power scales weakly with $\tau_0^{-1/2}$. However, above $\tau_0 \sim 50$ which is around the transition between Thomson and KN regime the power scales closer to $\tau_0^{-0.9}$. This being steeper than the scaling of $\tilde{D}_\mathrm{max}^{\hat{r}}$ suggests that the gap size decreases with increasing $\tau_0$ as well.

\subsection{Dependence on Resolution}
\label{sec:resolution}

We discovered that with increasing $\Bemin$, numerical simulations become more and more challenging. Figure \ref{fig:resolution} shows light curve comparisons between simulations with identical parameters, but different numbers of particles per cell, or different resolutions, at $\Bemin = 10^{4}$. With a smaller number of particles per cell or lower resolution, the simulation eventually settles to a quasi-steady state not unlike what was reported by \citet{2018A&A...616A.184L}. However, with more particles per cell or higher resolution, we observe the same kind of quasi-periodic opening and screening of the gap as shown in Section~\ref{sec:gap}. Generally we find that when the system settles to a quasi-steady state, not only is the dissipation much less than the peak of the cyclic solution, the average power is also about a a factor of $\sim 1.5$ less than the average power of the cyclic solution.

This resolution dependent behavior mainly happens with high $\tilde{B}$ and $\Bemin$, where large electrostatic oscillations can accelerate particles close to $\gamma\sim 1/\temin$, thus sustain pair creation without forming a time-dependent gap. To minimize this numerical effect, one needs both high spatial resolution and large number of particles per cell. The spatial grid should ideally resolve the plasma skin depth $\lambda_p$, otherwise the plasma will be heated to Lorentz factors $\gamma_T \sim (\Delta x / \lambda_p)^2$, at which point the hot plasma skin depth can be resolved. If this energy is already capable of producing pairs, we will simply have a hot thermal plasma in pair creation equilibrium. Small number of particles per cell will lead to high discrete particle noises in the simulation, which also translates to larger plasma oscillations, broadening the spread of distribution function in momentum space. If this allows more particles to reach pair creation energies then again a quasi-steady state can be achieved. We believe this is the main reason CY18 and \citet{2018A&A...616A.184L} found qualitatively different solutions to the same problem. Both were using a similar numerical resolution, but \citet{2018A&A...616A.184L} were attempting to use parameters corresponding to $\Bemin \sim 5\times 10^6$ while we found that $\Bemin = 10^4$ is already very challenging for the typical numerical resolution used in this kind of simulations. For this reason, we also expect that the gap power we see in our highest $\Bemin$ simulations may not be the most reliable (e.g.\ the last point in the left panel of Figure~\ref{fig:scaling}, where $\Bemin = 3\times 10^4$), as numerical effects reduce the amount of $E_{\parallel}$ generated by the system.

\section{Implications for M87}
\label{sec:m87}

M87, as a low luminosity AGN, is one of the most promising candidates where this process can operate. One can obtain a reasonable estimate for the magnetic field from the estimated jet power, $B_0\sim 200\,\mathrm{G}$ \citep[e.g.,][]{2011ApJ...730..123L,2015ApJ...809...97B}, which translates to $\tilde{B}_0 \sim 1.2\times 10^{14}$. However, the parameters for the soft photon field in the vincinity of the central black hole is a lot more uncertain. The SED for the radio flux seems to peak around 1.2 meV, which translates to $\temin \sim 2\times 10^{-9}$, but the soft photon spectrum near the horizon may well be very different from what is observed. Assuming all photons are coming from near the horizon, the upper bound for photon energy density is $u_s \sim 0.1\,\mathrm{erg/cm^3}$ which implies that $\tau_0 \sim 5\times 10^3$. But since most photons are produced in the disk, $u_s$ near the horizon could be orders of magnitude lower.

In CY18 we found that if we take $\tau_{0}\sim 5\times 10^{3}$ then the gap power in the Thomson regime would be $L\sim 3\times 10^{39}\,\mathrm{erg\,s^{-1}}$. If we assume much lower fiducial optical depth, e.g. $\tau_0 \sim 10$ as in our simulations, and simply scale $L/L_{0}$ according to Figure~\ref{fig:scaling} to realistic value of $\Bemin \sim 2.4\times 10^5$, we can expect the power dissipated in the gap to be $L \sim 10^{-3}L_0$. The jet power of M87 is usually estimated to be $L_0 \sim 10^{43}\text{--}10^{44}\,\mathrm{erg\,s^{-1}}$, so our estimated gap power should be around $L_\mathrm{gap} \sim 10^{40}\text{--}10^{41}\,\mathrm{erg\,s^{-1}}$, somewhat smaller than, but not entirely inconsistent with the reported power of the VHE flares from the same object \citep[e.g.][]{2012ApJ...746..151A}.

We expect the spectrum from the gap to be similar to what is shown in Figure~\ref{fig:spectra}, which is a power law extending up to $\sim 0.1/\temin = 5\times 10^7$, followed by an exponential cutoff. This translates to energy up to around $25\,\mathrm{TeV}$, also consistent with the observed flare spectrum. The exact slope of the spectrum will likely be subject to further modification due to scattering with the ambient soft photon field and radiation from the secondary pairs.

\section{Discussion}
\label{sec:discussion}

We presented results from first-principle 1D GR PIC simulations of the pair discharge process in the vincinity of supermassive black holes. In general we observe a highly time-dependent solution where an electric gap opens quasi-periodically near the null surface. The variability time scale of this activity is usually on the order of several $r_{g}/c$, and the outgoing photon spectrum forms a power law with index that depends on the gap activity.

In this work we ignored synchrotron and curvature radiation. In reality, when $\tau$ is order unity, photon free path is significant compared to the macroscopic length scale. As they convert to a pair, the particles should have a relatively large pitch angle with respect to the magnetic field. It will be damped quickly through synchrotron radiation, which should dissipate a sizable fraction of the particle energy and produce detectable radiation. The highest energy particles that have $\gamma \sim 10/\temin$ can approach curvature radiation reaction limit, and curvature loss can be a significant energy sink as well. We plan to carry out a more complete spectral analysis in the future, including the effect of synchrotron and curvature radiation in this framework, which could produce a unified spectrum from X-ray to VHE gamma rays out of the pair producing gaps.

In this paper we considered $\gamma$-$\gamma$ collision to be the only pair creation mechanism. As pointed out by \citet{2019arXiv190703175P}, when particles are in the deep KN regime, triplet pair production may play a role in the screening of the gap. Due to the highly uncertain nature of the soft photon field in the vincinity of the supermassive black hole of M87, there is a parameter regime where triplet pair production could be the main mechanism that regulates the gap physics. This can potentially further increase the gap luminosity and could be powering the strongest flares that we see up to $L\sim 10^{42}\,\mathrm{erg\,s^{-1}}$. A radiation module that handles triplet pair creation is already implemented in \emph{Aperture} and we expect to carry out a separate study for the importance of this process in M87, and whether this produces any observable signatures on the spectrum.

This paper studied the gap microphysics in 1D, whereas in many runs the gap evolves to have a size comparable to $r_g$, at which point the 1D approximation is not completely valid. Working 2D axisymmetric GR PIC codes now exist \citep{2019PhRvL.122c5101P}. It is an important and interesting question whether the results presented in this paper are robust when the 1D approximation is relaxed. However, we caution any future attempt at solving this problem in 2D full GR that resolution is important. Significant rescaling and some abstraction of the radiative physics must be employed, and extreme care must be taken in order to draw any meaningful comparisons with observations.

\acknowledgments
We thank Amir Levinson, Frank Rieger, Kouichi Hirotani, Zhen Pan, and Huan Yang for helpful discussion. The code \emph{Aperture} used in this work can be found at https://github.com/fizban007/Aperture3.git. AC acknowledges support from NASA grant NNX15AM30G. YY acknowledges support from the Lyman Spitzer, Jr.~Postdoctoral Fellowship. We also gratefully acknowledge the support of NVIDIA Corporation with the donation of the Quadro P6000 GPU used for this research.

\bibliographystyle{aasjournal}


\appendix
\twocolumngrid

\section{Background force-free configuration}\label{sec:force-free}
\begin{figure}
    \centering
    \includegraphics[width=0.9\columnwidth]{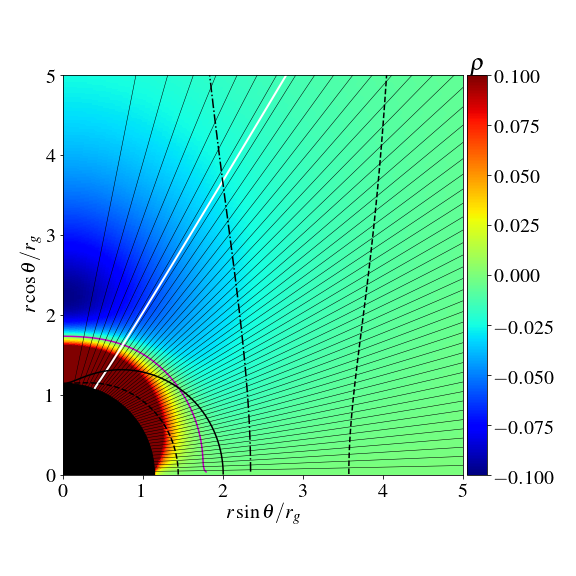}
    \caption{Force-free solution of a monopolar magnetosphere around a Kerr black hole with spin $a=0.99$. The thin black lines show the flux surfaces with constant $\psi$, and the white line indicates the specific flux tube we used as the background for our kinetic simulations presented in this paper. The thick black line shows the ergosphere; the dashed black lines are the light surfaces; the dot-dashed black line indicates the stagnation surface. The color on the plot shows the charge density $\rho$ as measured by ZAMO, and the magenta line marks the null surface where $\rho=0$.}
    \label{fig:FFsolution}
\end{figure}

We consider a split-monopole field configuration in Kerr spacetime. The axisymmetric, steady-state, force-free solution is obtained using a numerical method described in \citet{2019MNRAS.484.4920Y}. Figure \ref{fig:FFsolution} shows one example with $a=0.99$, which we used in this paper.

The background solution is specified by the magnetic flux $\psi$ and two flux functions: $\Omega(\psi)$---angular velocity of the field line, and $B_T(\psi)$---current flowing along the flux surfaces. The components of the magnetic field can be obtained as
\begin{align}
    B^{\phi}=\frac{1}{\sqrt{\gamma}}\frac{B_T g_{rr}g_{\theta\theta}}{\sqrt{-g}},\ 
    B^r=\frac{\partial_{\theta}\psi}{\sqrt{\gamma}},\ 
    B^{\theta}=-\frac{\partial_{r}\psi}{\sqrt{\gamma}}.
\end{align}
The background charge and current densities satisfy
\begin{equation}
    j_{\rm ff}^{\phi}=\Omega \rho_{\rm ff}+\frac{B^{\phi}}{B^r}j_{\rm ff}^r,
\end{equation}
and the corresponding components are
\begin{align}
    \rho&=\alpha J^0=-\frac{1}{\sqrt{\gamma}}\partial_A\left(\frac{\sqrt{-g}}{\Delta \sin^2\theta}g^{AB}(\Omega-\omega)\partial_B\psi\right),\nonumber\\ & \text{where } A,B=r,\theta,\\
    j^r&=\alpha J^r=\frac{B_{T,\theta}}{\sqrt{\gamma}},\quad
    j^{\theta}=\alpha J^{\theta}=-\frac{B_{T,r}}{\sqrt{\gamma}}.
\end{align}

\end{document}